\documentclass[conference]{IEEEtran}
\IEEEoverridecommandlockouts
\usepackage{cite}
\usepackage{amsmath,amssymb,amsfonts}
\usepackage{algorithmic}
\usepackage{graphicx}
\usepackage{textcomp}
\usepackage{xcolor}
\usepackage{caption}
\usepackage{subcaption}
\usepackage{xcolor}
\usepackage{hyperref}
\renewcommand{\baselinestretch}{0.95}
\def\BibTeX{{\rm B\kern-.05em{\sc i\kern-.025em b}\kern-.08em
    T\kern-.1667em\lower.7ex\hbox{E}\kern-.125emX}}

\usepackage{tikz}
\newcommand\copyrighttext{%
  \footnotesize 979-8-3315-8745-1/25/\$31.00~\copyright~2025 IEEE. Personal use permitted. Published in: 2025 IEEE Asilomar. DOI: 10.1109/IEEECONF67917.2025.11443710}
\newcommand\copyrightnotice{%
\begin{tikzpicture}[remember picture,overlay]
\node[anchor=south, yshift=15pt] at (current page.south) {\copyrighttext};
\end{tikzpicture}%
}

\begin{document}

\title{Elevation- and Tilt-Aware Shadow Fading Correlation Modeling for UAV Communications\\ 
\thanks{We use the publicly available field measurement data and the preprocessing scripts~\cite{IEEEDataPort_3, masrur2025collection} published by the NSF Aerial Experimentation and Research Platform on Advanced Wireless (AERPAW)~\cite{9061157} platform. This work is supported in part by the INL Laboratory Directed Research Development (LDRD) Program under BMC No. 264247, Release No. 26 on BEA’s Prime Contract No. DE-AC07-05ID14517.}
\thanks{The code, along with the dataset and reproducible results, is available at \url{https://github.com/MPACT-Lab/shadow-fading-ang-corr}.} 
}

\author{\IEEEauthorblockN{Mushfiqur Rahman$^{\ast}$, \.{I}smail G\"{u}ven\c{c}$^{\ast}$, Mihail Sichitiu$^{\ast}$, Jason A. Abrahamson$^{\dagger}$,\\
Bryton J. Petersen$^{\dagger}$, Amitabh Mishra$^{\dagger}$, and Arupjyoti Bhuyan$^{\dagger}$}
\IEEEauthorblockA{$^{\ast}$\textit{Department of Electrical and Computer Engineering, North Carolina State University, Raleigh, NC, USA} \\
$^{\dagger}$\textit{Idaho National Laboratory, Idaho Falls, ID, USA}\\
{\tt\small \{mrahman7, iguvenc, mlsichit\}@ncsu.edu}\\
{\tt\small \{jason.abrahamson, Bryton.Petersen, Amitabh.Mishra, arupjyoti.bhuyan\}@inl.gov}\\
}
}


\maketitle
\copyrightnotice

\begin{abstract}
Future wireless networks demand a more accurate understanding of channel behavior to enable efficient communication with reduced interference. Uncrewed Aerial Vehicles (UAVs) are poised to play an integral role in these networks, offering versatile applications and flexible deployment options. However, accurately characterizing the shadow fading (SF) behavior in UAV communications remains a challenge.
Traditional SF correlation models rely on spatial distance and neglect the UAV's 3D orientation and elevation angle. Yet even slight variations in pitch angle (5–10\(^\circ\)) can significantly affect the signal strength observed by a UAV.
In this study, we investigate the impact of UAV pitch and elevation geometry on SF and propose an elevation- and tilt-aware spatial correlation model. We use a real-world fixed-altitude UAV measurement dataset collected in a rural environment at 3.32 GHz with a 125 kHz bandwidth. Results show that a 10\(^\circ\) tilt-angle separation and a 20\(^\circ\) elevation-angle separation can reduce the SF correlation by up to 15\% and 40\%, respectively.
In addition, integrating the proposed correlation model into the ordinary Kriging (OK) framework for signal strength prediction yields an approximate 1.5 dB improvement in median RMSE relative to the traditional correlation model that ignores UAV orientation and elevation.
\end{abstract}


\section{Introduction}
The growth and influence of aerial devices in wireless systems are rapidly increasing~\cite{9061157}. Aerial platforms, including uncrewed aerial vehicles (UAVs), low-altitude platforms (LAPs), high-altitude platforms (HAPs), and satellites, are expected to play a pivotal role in the development of future 6G technologies. UAV-based base stations (UAV-BSs), in particular, are anticipated to become a key component of next-generation wireless networks~\cite{9061157}. Consequently, precise path loss modeling is crucial for ensuring reliable, efficient, and uninterrupted communication in these evolving systems.
Unlike terrestrial devices, such as mobile phones, Internet of Things (IoT) devices, and radars, which typically operate at low altitudes and along a horizontal plane, aerial platforms can traverse a wide range of altitudes and 3D orientations, including roll, yaw, and pitch angles. This expanded operational space introduces increased complexity in establishing optimal communication links, as the performance of aerial devices is highly sensitive to both their altitude and 3D orientation~\cite{9061157, romero2020aerial}. In particular, airframe shadowing, which depends on the airframe and antenna mounting design, can significantly affect the received signal power when the line-of-sight (LoS) path is diffracted or blocked by the airframe~\cite{sun2017air, ge2023pathloss, ni2024path}. 

The correlation of shadow fading (SF) plays a crucial role in various applications, including received signal strength prediction, radio environment map (REM) construction~\cite{sasaki2024improving}, power control~\cite{cui2020multi}, and cognitive radio networks (CRNs)~\cite{shen2019uav}. For example, Kriging-based REM generation decomposes the received power into path loss (PL) and SF components, with the SF predicted using a shadowing correlation matrix~\cite{maeng2023kriging}. SF, in general, refers to the signal strength fluctuations that occur gradually over an area and remain relatively consistent over time. This gradual nature is captured by the correlation between SF values at different locations. The correlations are learned from measurement data. UAVs, for instance, can transmit signal strength measurements to a central control station, which can then be shared with other UAVs as a shadowing-correlation model to ease channel prediction, especially when similar obstacles contribute to shadowing. 

Traditionally, shadowing correlation is modeled as a function of horizontal and vertical distances between receiver locations, with the assumption of stationary covariance~\cite{maeng2023kriging}. However, several studies have demonstrated that this stationarity assumption does not hold for SF, as it is influenced by other factors such as antenna tilt and elevation angle~\cite{kifle2013impact, seah2024study, sharma2018study}. UAV-based measurement campaigns have further revealed significant variations in SF distributions across different elevation angles and UAV 3D orientations, primarily due to shadowing effects caused by the UAV's body~\cite{badi2020experimentally}. Despite these findings, existing literature has not numerically characterized how SF correlation varies with the UAV's 3D pose or elevation angle.
In this work, we model the angular variations of SF correlation using an exponential kernel that explicitly accounts for elevation- or orientation-induced changes. The resulting angular-dependent correlation functions are then combined with a conventional distance-dependent function. The main contributions of this paper are summarized as follows:
\begin{itemize}
\item We model the variation of SF across elevation angles by an exponentially decaying correlation function of the elevation-angle separation. Different decay constants are employed for increasing and decreasing elevation, capturing the asymmetry of the correlation behavior with respect to elevation changes.
\item We introduce the UAV tilt angle toward the transmitter as an indicator variable that captures SF variations induced by the UAV’s 3D pose, thereby reducing the three orientation components (roll, pitch, and yaw) to a single dominant factor for correlation modeling.
\item We use an exponential kernel to model SF correlation across tilt-angle separation. Analogous to the elevation case, we adopt a piecewise formulation with distinct decay constants for positive and negative tilt changes.
\item Using real UAV measurement data, we learn a joint correlation profile parameterized by \((\text{elevation}, \text{tilt})\) pairs, demonstrating that the elevation-dependent correlation decay is tilt-angle dependent and vice versa. 
\item We use the proposed correlation profile with the ordinary Kriging (OK) framework for signal strength prediction and show a significant improvement in accuracy. 
\end{itemize}
The rest of this paper is organized as follows. Section~\ref{sec:bg_and_prob_formulation} provides the background for this study. Section~\ref{sec:proposed_method} introduces the proposed SF correlation model. Section~\ref{sec:exp_results} presents the experimental results from real UAV measurements. Finally, Section~\ref{sec:conclution} concludes the paper.

\section{Background and Problem Formulation} 
\label{sec:bg_and_prob_formulation}
\subsection{System Model} 
\begin{figure}[!t]
\centerline{\includegraphics[width=0.96\linewidth,trim={11.0cm 2.5cm 7.4cm 5.5cm},clip]{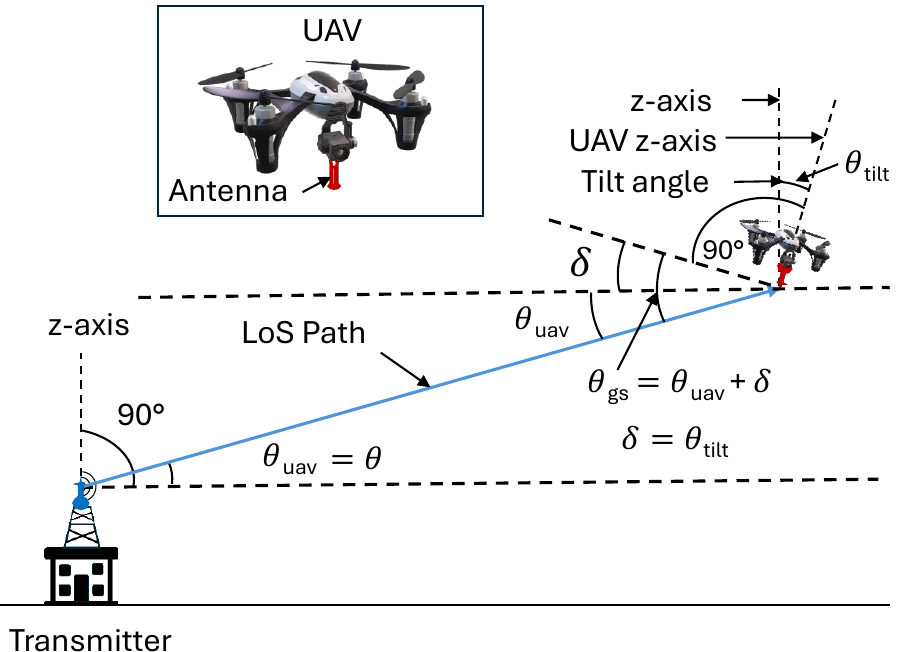}}
\caption{System model for UAV-based measurements.}
\label{fig:ststem_model}
\vspace{-5mm}
\end{figure}

The measurement system, illustrated in Fig.~\ref{fig:ststem_model}, consists of a UAV equipped with a receive antenna mounted beneath its fuselage. The UAV follows a fixed-altitude, arbitrary horizontal trajectory. At each sampling instant along this trajectory, the UAV records the reference signal received power (RSRP) \(z\), its 3D position \(\mathbf{s} \in \mathbb{R}^{3}\) given by latitude, longitude, and altitude, and its 3D orientation, represented by Z–Y–X Euler angles \( \alpha \), \( \beta \), and \( \gamma \), corresponding to yaw, pitch, and roll, respectively. The complete set of measurements acquired along the trajectory is denoted by \(\mathcal{D}\), which consists of \(N\) samples, each representing measurements taken over a short time interval.
In a post-processing stage, given the dataset \(\mathcal{D}\), we compute the elevation angle of the UAV as seen from the transmitter, denoted by \(\theta_\text{uav}\) or simply \(\theta\). Using the UAV’s 3D orientation \((\alpha, \beta, \gamma\)), we also compute the elevation angle of the ground station as seen from the UAV antenna~\cite{craig2018introduction}, denoted by \(\theta_\text{gs}\). The difference between the two elevation angles, \(\theta_\text{uav}\) and \(\theta_\text{gs}\), quantifies the angle by which the UAV’s vertical axis is tilted toward or away from the transmitter. We denote this elevation-angle difference by \(\delta\) and refer to it as the UAV tilt angle. In other words, \(\delta\) is identical to the UAV vertical tilt angle \(\theta_\text{tilt}\). The main notation used throughout the paper is summarized in Table~\ref{tab:variable_symbol_list}. 
\renewcommand{\arraystretch}{1.1}
\begin{table*}[!t]
\centering
\caption{List of main symbols, units, and data gathering stage.}
\label{tab:variable_symbol_list}
\begin{tabular}{c l c l l}
\hline
\textbf{Symbol} & \multicolumn{1}{c}{\textbf{Description}} & \textbf{Unit} & \multicolumn{1}{c}{\textbf{Stage}} & \multicolumn{1}{c}{\textbf{Remarks}} \\ \hline
\(z\)                   & Reference Signal Received Power (RSRP)                     & dBm & UAV measurement      & Standard quantity            \\
\(\mathbf{s}\)          & UAV 3D position (latitude, longitude, altitude)       & (deg, deg, m)      & UAV measurement & Standard quantity      \\ 
\(\alpha,\beta,\gamma\) & Roll, pitch, yaw Euler angles                       & deg     & UAV measurement & Standard quantity   \\ \hline
\(w\)                   & Shadow fading (SF) component                        & dB    & Post-processing    & Standard concept           \\
\(\theta\)              & UAV elevation angle                                 & deg    & Post-processing   & Standard quantity           \\
\(\delta\)              & UAV vertical tilt angle                             & deg    & Post-processing   & Defined in this paper       \\
\(d_{\text{2D}}(\mathbf{s}_i,\mathbf{s}_j)\) 
                        & Horizontal distance between locations \(i\) and \(j\) & m     & Post-processing  & Standard quantity       \\ \hline
\(\mu\)                 & Mean of shadow fading                               & dB        & Correlation modeling  & Standard quantity          \\
\(\sigma^{2}\)          & Variance of shadow fading                           & dB\(^2\)   & Correlation modeling & Standard quantity        \\
\(R(\mathbf{s}_i,\mathbf{s}_j)\) 
                        & Spatial correlation function                        & --        & Correlation modeling & Standard concept         \\
\(R_\text{ang}(\theta_i,\theta_j,\delta_i,\delta_j)\) 
                        & Angular correlation function                        & --       & Correlation modeling  & Defined in this paper     \\ \hline
\end{tabular}
\vspace{-4mm}
\end{table*}

\subsection{Problem Statement}
Consider an extensive UAV measurement dataset \(\mathcal{D}'\) consisting of \(N'\) samples, where each sample includes the measured RSRP \(z\), the UAV’s 3D position \(\mathbf{s}\), and its 3D orientation angles \(\alpha\), \(\beta\), and \(\gamma\). The goal is to learn the SF characteristics of the environment from this dataset, which may include the SF mean \(\mu\), variance \(\sigma^2\), and the correlation of SF between two points \(R(\mathbf{s}_i,\mathbf{s}_j)\). The effectiveness of SF learning is evaluated using a separate test dataset \(\mathcal{D}\) with \(N\) samples. 
For testing, a subset of \(M\) samples from \(\mathcal{D}\) (with \(M \ll N'\)) is selected as tuning samples. These \(M\) samples are used to predict \(z\) values for the remaining \(N-M\) samples in \(\mathcal{D}\), given their corresponding values for \(\mathbf{s}\), \(\alpha\), \(\beta\), and \(\gamma\). This scenario is analogous to having a limited number of \(M\) sensors deployed across the region or receiving measurement reports from \(M\) users, from which we aim to predict values at other locations and orientations. The prediction leverages the learned SF characteristics from \(\mathcal{D}'\), with the transmitter’s 3D position, transmit power, and antenna pattern assumed to be known.

\subsection{Baseline Approach}
OK is widely used for RSRP estimation at unmeasured locations and is adopted as the baseline interpolation method in this study. We apply OK to the SF component after decomposing the measured RSRP into a deterministic PL term and a random SF term as follows:
\begin{equation}
z(i) = \text{PL}_{\text{tr}}\big(\mathbf{s}_i,\mathbf{s}_\text{Tx},G_\text{uav},G_\text{Tx},P_\text{Tx},f_0,\Gamma\big) + w(i),
\label{eq:pl_sf_decomposition}
\end{equation}
where \(z(i)\), \(w(i)\), and \(\mathbf{s}_i\) denote the RSRP, SF, and 3D UAV position corresponding to the \(i\)-th sample from the dataset \(\mathcal{D}\), respectively; \(\text{PL}_{\text{tr}}(\cdot)\) is the two-ray path loss model used for RSRP estimation~\cite{maeng2023kriging}; \(\mathbf{s}_\text{Tx}\) is the 3D position of the transmitter; \(G_\text{uav}\) and \(G_\text{Tx}\) are the corresponding antenna gains; \(P_\text{Tx}\) is the transmit power; \(f_0\) is the carrier frequency; and \(\Gamma\) denotes the ground reflection coefficient.

Since the PL component at an arbitrary prediction location can be evaluated directly using the same two-ray model \(\text{PL}_{\text{tr}}(\cdot)\), the remaining uncertainty resides in the SF term \(w\). Let \(\mathbf{s}_0 \in \mathbb{R}^{3}\) denote an unmeasured 3D location, and let \(w(0) \in \mathbb{R}\) denote the corresponding (unknown) SF to be estimated. Consider \(M\) training samples with SF values \(\mathbf{w} = [w(1),\dots,w(M)]^\mathsf{T} \in \mathbb{R}^{M}\) and associated 3D positions \(\mathbf{s}_1,\dots,\mathbf{s}_M\). OK forms a linear predictor comprising all training samples as follows:
\begin{equation}
\hat{w}(0) = \sum_{i=1}^{M} \lambda_i\, w(i) = \boldsymbol{\lambda}^\mathsf{T}\mathbf{w},
\label{eq:ok_predictor}
\end{equation}
where \(\boldsymbol{\lambda} = [\lambda_1,\dots,\lambda_M]^\mathsf{T} \in \mathbb{R}^{M}\) are the Kriging weights. The weights \(\boldsymbol{\lambda}\) are determined by solving the standard OK linear system as follows:
\begin{equation}
\begin{bmatrix}
\mathbf{C} & \mathbf{1} \\
\mathbf{1}^\mathsf{T} & 0
\end{bmatrix}
\begin{bmatrix}
\boldsymbol{\lambda} \\[4pt] \nu
\end{bmatrix}
=
\begin{bmatrix}
\mathbf{c}_0 \\[4pt] 1
\end{bmatrix},
\label{eq:ok_system}
\end{equation}
where \(\mathbf{C} \in \mathbb{R}^{M\times M}\) is the covariance matrix with entries
\(
C_{ij} = C(\mathbf{s}_i,\mathbf{s}_j),
\)
\(\mathbf{c}_0 = [C(\mathbf{s}_1,\mathbf{s}_0),\dots,C(\mathbf{s}_M,\mathbf{s}_0)]^\mathsf{T}\) is the covariance vector between the training samples and the prediction sample, \(\mathbf{1} = [1,\dots,1]^\mathsf{T} \in \mathbb{R}^{M}\) is the all-ones vector, and \(\nu\) is the Lagrange multiplier. In the conventional approach, the SF field is assumed to be second-order stationary, with the covariance function modeled as:
\begin{equation}
C(\mathbf{s}_i,\mathbf{s}_j) = \sigma^2 R(\mathbf{s}_i,\mathbf{s}_j),
\end{equation}
where \(\sigma^2\) is the SF variance and \(R(\mathbf{s}_i,\mathbf{s}_j)\) is the spatial correlation function between locations \(\mathbf{s}_i\) and \(\mathbf{s}_j\). In this study, the baseline (conventional) correlation model adopts a double-exponential decaying model (DEDM)~\cite{algans2002experimental} as follows:
\begin{equation}
R(\mathbf{s}_i,\mathbf{s}_j) = a e^{-p_1 d_{\text{2D}}(\mathbf{s}_i,\mathbf{s}_j)} + (1-a)e^{-p_2 d_{\text{2D}}(\mathbf{s}_i,\mathbf{s}_j)},
\label{eq:dedm}
\end{equation}
where \(d_{\text{2D}}(\mathbf{s}_i,\mathbf{s}_j)\) denotes the horizontal distance between \(\mathbf{s}_i\) and \(\mathbf{s}_j\), and \(a \in [0,1]\), \(p_1 > 0\), and \(p_2 > 0\) are correlation-model parameters estimated from the dataset \(\mathcal{D}'\). The model in~\eqref{eq:dedm} is a purely distance-based, stationary correlation function that depends solely on the horizontal separation between points, without accounting for variations in UAV elevation or orientation. In contrast, the proposed approach relaxes this stationarity assumption by incorporating elevation- and tilt-dependent correlation factors. 

\section{Proposed Shadow Fading Correlation Model}
\label{sec:proposed_method}
We propose the following separable correlation function, which separates the spatial and angular dependencies:
\begin{equation}
\hat{R}(\mathbf{s}_i,\mathbf{s}_j,\theta_i,\theta_j,\delta_i,\delta_j)
= R(\mathbf{s}_i,\mathbf{s}_j)\, R_\text{ang}(\theta_i,\theta_j,\delta_i,\delta_j),
\label{eq:proposed_full_r}
\end{equation}
where \(\hat{R}(\cdot)\) denotes the proposed correlation function, \(R(\cdot)\) is the distance-based spatial correlation function given in~\eqref{eq:dedm}, and \(R_\text{ang}(\cdot)\) captures the correlation contribution due to elevation angles \(\theta_i,\theta_j\) and tilt angles \(\delta_i,\delta_j\). In the conventional baseline, the angular contribution is implicitly ignored by setting \(R_\text{ang}(\cdot) \equiv 1\). To further disentangle the effects of tilt and elevation, we factorize the angular correlation into two components as follows:
\begin{equation}
R_\text{ang}(\theta_i,\theta_j,\delta_i,\delta_j)
= R_\text{tlt}(\delta_i,\delta_j,\theta_i)\,
  R_\text{elv}(\theta_i,\theta_j,\delta_i),
\end{equation}
where \(R_\text{tlt}(\delta_i,\delta_j,\theta_i)\) accounts for the correlation variation induced by tilt-angle separation at a fixed elevation angle, and \(R_\text{elv}(\theta_i,\theta_j,\delta_i)\) captures the impact of elevation-angle separation at a fixed tilt angle. In the following subsections, we explain how these two components are empirically estimated from the dataset \(\mathcal{D}'\) and fitted using exponential kernels.

\subsection{Tilt-angle Separation at Fixed Elevation Angle $R_\text{tlt}(\cdot)$}
Consider SF observations under two tilt configurations \(\delta_i\) and \(\delta_j\) at a fixed elevation angle \(\theta_i\). For simplicity, we assume equal numbers of observations \(n\). If the number of observations differs between the two configurations, resampling can be used to match the sample sizes \(n\) based on the empirical SF distributions. Let \(\mathbf{w}_{ii} = [w_{ii}(1),\dots,w_{ii}(n)]\) denote \(n\) SF samples measured at \((\theta,\delta) = (\theta_i,\delta_i)\), and let \(\mathbf{w}_{ij} = [w_{ij}(1),\dots,w_{ij}(n)]\) denote \(n\) SF samples measured at \((\theta,\delta) = (\theta_i,\delta_j)\). Without loss of generality, assume that both \(\mathbf{w}_{ii}\) and \(\mathbf{w}_{ij}\) are sorted in ascending order.

We now introduce a simple tilt-induced transformation model. Given an SF realization \(w_{ii}(k)\) at \((\theta_i,\delta_i)\), we define the corresponding “converted” SF value at \((\theta_i,\delta_j)\) as follows:
\begin{equation}
w_{ii\to ij}(k) = f_{\delta_i \to \delta_j \mid \theta_i}\big(w_{ii}(k)\big)
= w_{ii}(k) + \Delta w,
\label{eq:tilt_shift_model}
\end{equation}
where \(w_{ii\to ij}(k)\) denotes the SF that would be observed if only the tilt angle were changed from \(\delta_i\) to \(\delta_j\), while all other measurement conditions remained unchanged, and \(\Delta w\) is a tilt-induced offset. In model~\eqref{eq:tilt_shift_model}, the term \(w_{ii}(k)\) is an SF component, and \(\Delta w\) is a constant for the conversion scenario (\(\delta_i \to \delta_j \mid \theta_i\)). Under this model, \(f_{\delta_i \to \delta_j \mid \theta_i}(\cdot)\) is a strictly increasing function of \(w_{ii}(k)\); hence, if \(\mathbf{w}_{ii}\) is sorted in ascending order, then the transformed vector \(f_{\delta_i \to \delta_j \mid \theta_i}(\mathbf{w}_{ii})\) is also sorted in ascending order. If the dataset \(\mathcal{D}'\) is \emph{balanced}, meaning that for each SF sample in \(\mathbf{w}_{ii}\), there exists a corresponding SF sample in \(\mathbf{w}_{ij}\), then based on the “ascending” order correspondence, we can express:
\begin{equation}
w_{ii\to ij}(k) \approx w_{ij}(k), \quad k = 1,\dots,n.
\label{eq:tilt_equivalence}
\end{equation}
Under this balancing assumption, the identification in~\eqref{eq:tilt_equivalence} provides the basis for estimating an empirical tilt-dependent correlation function \(R_\text{tlt}^{\text{emp}}(\cdot)\) as follows:
{
\small
\begin{equation}
R_\text{tlt}^{\text{emp}}(\delta_i,\delta_j,\theta_i)
=
\frac{\sum_{k=1}^{n}\big(w_{ii}(k)-\mu\big)\big(w_{ij}(k)-\mu\big)}
     {\sqrt{\sum_{k=1}^{n}\big(w_{ii}(k)-\mu\big)^2}\,
      \sqrt{\sum_{k=1}^{n}\big(w_{ij}(k)-\mu\big)^2}},
\label{eq:emp_tilt_corr}
\end{equation}
}
where \(\mu\) denotes the SF mean learned from \(\mathcal{D}'\). Equation~\eqref{eq:emp_tilt_corr} is simply the sample correlation coefficient between the SF vectors \(\mathbf{w}_{ii}\) and \(\mathbf{w}_{ij}\) at elevation angle \(\theta_i\).

We then model the tilt-dependent correlation as a function of the tilt-angle separation \(|\delta_i-\delta_j|\) using an exponential kernel. To allow for asymmetric sensitivity with respect to tilt variation, we adopt the following piecewise form:
\begin{equation}
R_\text{tlt}(\delta_i,\delta_j,\theta_i) =
\begin{cases}
\exp\!\big(-|\delta_i-\delta_j|/q_1\big), & \delta_j \ge \delta_i \\[4pt]
\exp\!\big(-|\delta_i-\delta_j|/q_2\big), & \delta_j < \delta_i
\end{cases},
\label{eq:tilt_kernel}
\end{equation}
where \(q_1 > 0\) and \(q_2 > 0\) are kernel decay parameters learned from data (e.g., by fitting~\eqref{eq:tilt_kernel} to the empirical values in~\eqref{eq:emp_tilt_corr} as a function of \(|\delta_i-\delta_j|\)). The piecewise structure reflects the fact that, unlike horizontal separation, changes in tilt angle in opposite directions do not have symmetric propagation effects: one direction may block the LoS path and enhance diffraction, whereas the other may reduce blockage and partially restore the LoS component.

\subsection{Elevation-angle Separation at Fixed Tilt Angle $R_\text{elv}(\cdot)$}
We consider SF observation subsets for two elevation angles, \(\theta_i\) and \(\theta_j\), under a fixed tilt angle \(\delta_i\). As with the tilt-angle separation case, we assume, without loss of generality, that both subsets contain the same number of observations and are sorted in ascending order. Let \(\mathbf{w}_{ii} = [w_{ii}(1), \dots, w_{ii}(n)]\) represent \(n\) SF samples measured at \((\theta, \delta) = (\theta_i, \delta_i)\), and let \(\mathbf{w}_{ji} = [w_{ji}(1), \dots, w_{ji}(n)]\) represent \(n\) SF samples measured at \((\theta, \delta) = (\theta_j, \delta_i)\). A key distinction from the tilt-angle separation scenario is that changing the elevation angle inherently results in spatial (horizontal) separation of the UAV. Thus, unlike tilt separation, elevation separation cannot be realized without corresponding changes in spatial location.

We model the transformation of an SF realization \(w_{ii}(k)\) from \(\theta_i\) to the corresponding SF realization at \(\theta_j\) as:
\begin{equation}
w_{ii \to ji}(k) = f_{\theta_i \to \theta_j \mid \delta_i}\big(f_{\text{d2D}}(w_{ii}(k))\big) = f_{\text{d2D}}\big(w_{ii}(k)\big) + \Delta w,
\label{eq:elev_shift_model}
\end{equation}
where \(w_{ii \to ji}(k)\) represents the predicted SF realization at elevation \(\theta_j\), \(f_{\text{d2D}}(\cdot)\) is a mapping that accounts for spatial relocation based on the stationary covariance model in~\eqref{eq:dedm}, meaning the empirical distribution of \(f_{\text{d2D}}(w_{ii}(k))\) is identical to that of \(w_{ii}(k)\), and \(\Delta w\) is the elevation-induced offset. The elevation-dependent correlation term \(R_{\text{elv}}(\cdot)\) is defined as the correlation between \(f_{\text{d2D}}(\mathbf{w}_{ii})\) and \(\mathbf{w}_{ji}\).

Assuming a balanced dataset, where each sample in \(\mathbf{w}_{ii}\) corresponds to a matching sample in \(\mathbf{w}_{ji}\), we can write:
\begin{equation}
w_{ii \to ji}(k) \approx w_{ji}(k'), \,\, k \in [1, \dots, n], \,\, k' \in [1, \dots, n],
\label{eq:elevation_equivalence}
\end{equation}
where \(k\) and \(k'\) denote the corresponding indices between \(\mathbf{w}_{ii \to ji}\) and \(\mathbf{w}_{ji}\). By substituting~\eqref{eq:elevation_equivalence} into~\eqref{eq:elev_shift_model}, we obtain:
\begin{equation}
w_{ji}(k') = f_{\text{d2D}}\big(w_{ii}(k)\big) + \Delta w.
\label{eq:elev_shift_model2}
\end{equation}
Again, \(R_{\text{elv}}(\cdot)\) is the correlation between \(f_{\text{d2D}}(\mathbf{w}_{ii})\) and \(\mathbf{w}_{ji}\). According to~\eqref{eq:elev_shift_model2}, these two variables differ only by an offset, meaning their one-to-one correspondence is simply their sorted samples, which can be obtained from their empirical distributions. As stated earlier, empirical distributions of \(f_{\text{d2D}}(\mathbf{w}_{ii})\) and \(\mathbf{w}_{ii}\) are the same. As a result, \(R_{\text{elv}}^{\text{emp}}(\theta_i, \theta_j, \delta_i)\) can be computed using the same sample-correlation formula as in~\eqref{eq:emp_tilt_corr}, replacing \(\mathbf{w}_{ij}\) with \(\mathbf{w}_{ji}\).

Finally, we model the correlation function based on the elevation-angle separation \(|\theta_i-\theta_j|\) as follows:
\begin{equation}
R_\text{elv}(\theta_i,\theta_j,\delta_i) =
\begin{cases}
\exp\!\big(-|\theta_i-\theta_j|/r_1\big), & \theta_j \ge \theta_i \\[4pt]
\exp\!\big(-|\theta_i-\theta_j|/r_2\big), & \theta_j < \theta_i
\end{cases},
\label{eq:elev_kernel}
\end{equation}
where \(r_1 > 0\) and \(r_2 > 0\) are decay parameters fitted to the empirical correlation values. Similar to the tilt case, the piecewise structure is used to account for the asymmetric effect of increasing versus decreasing elevation angle.

\section{Experimental Results}
\label{sec:exp_results}
\subsection{Datasets and Evaluation Criteria}
We evaluate the proposed approach using a publicly available UAV measurement dataset collected in a rural environment, where a UAV flies in the vicinity of an unmanned ground vehicle (UGV) carrying the signal transmitter~\cite{masrur2025collection, IEEEDataPort_3}. The UAV is equipped with a USRP B205 software-defined radio, RF front-ends, and an SA-1400-5900 antenna~\cite{sa_1400}. The transmitter, mounted on the UGV, also uses an SA-1400-5900 antenna at an approximate height of \(1.5\)~m. The carrier frequency is \(3.32\)~GHz with a signal bandwidth of \(125\)~kHz.
\begin{figure}[t!]
    \centering
    \begin{subfigure}{0.23\textwidth}
        \centering
        \includegraphics[width=\linewidth,trim={3.4cm 7.6cm 4.4cm 8.35cm},clip]{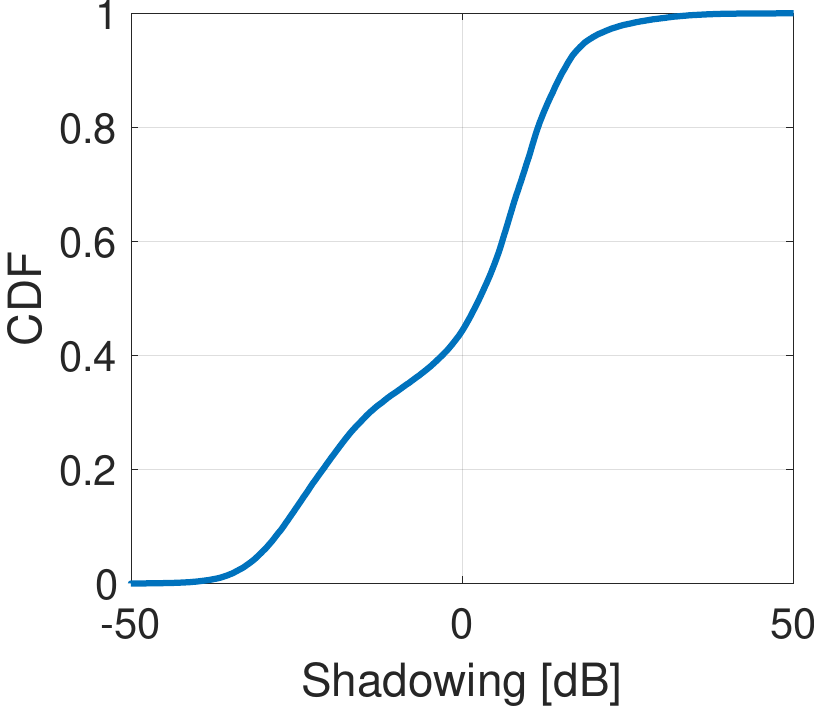}
        \caption{Unconditional}
        \label{fig:sub1}
    \end{subfigure}
    \begin{subfigure}{0.23\textwidth}
        \centering
        \includegraphics[width=\linewidth,trim={3.4cm 7.6cm 4.4cm 8.35cm},clip]{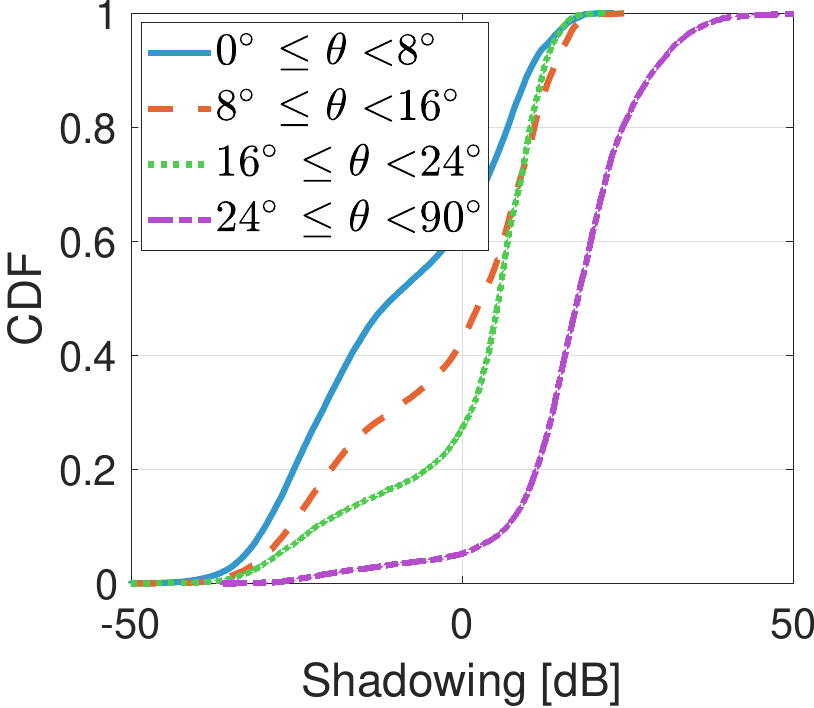}
        \caption{Elevation}
        \label{fig:sub1}
    \end{subfigure}
    \begin{subfigure}{0.23\textwidth}
        \centering
        \includegraphics[width=\linewidth,trim={3.4cm 7.6cm 4.4cm 8.15cm},clip]{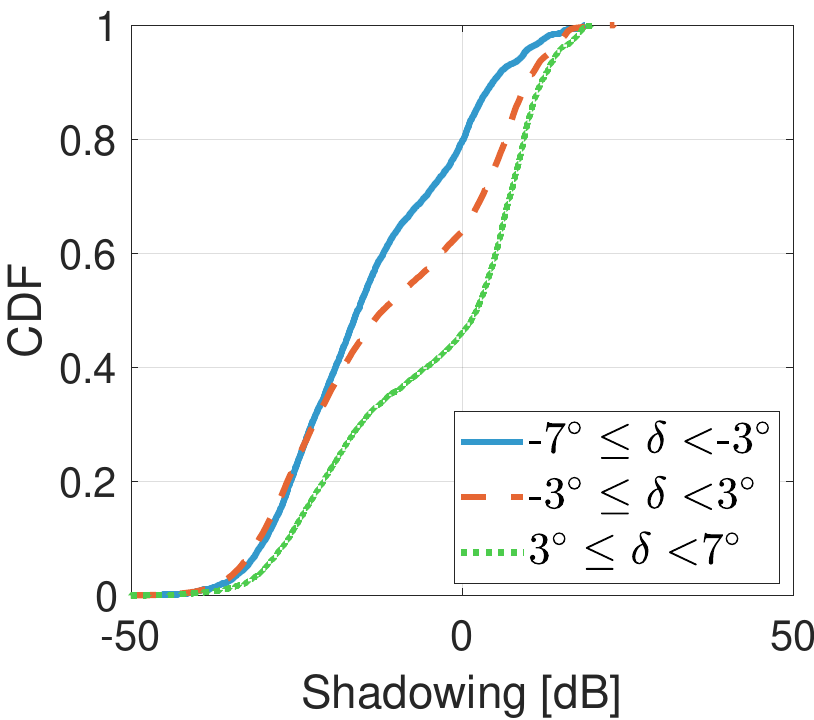}
        \caption{Tilt, \(0^\circ <\theta \leq 10^\circ\)}
        \label{fig:sub1}
    \end{subfigure}
    \begin{subfigure}{0.23\textwidth}
        \centering
        \includegraphics[width=\linewidth,trim={3.4cm 7.6cm 4.4cm 8.15cm},clip]{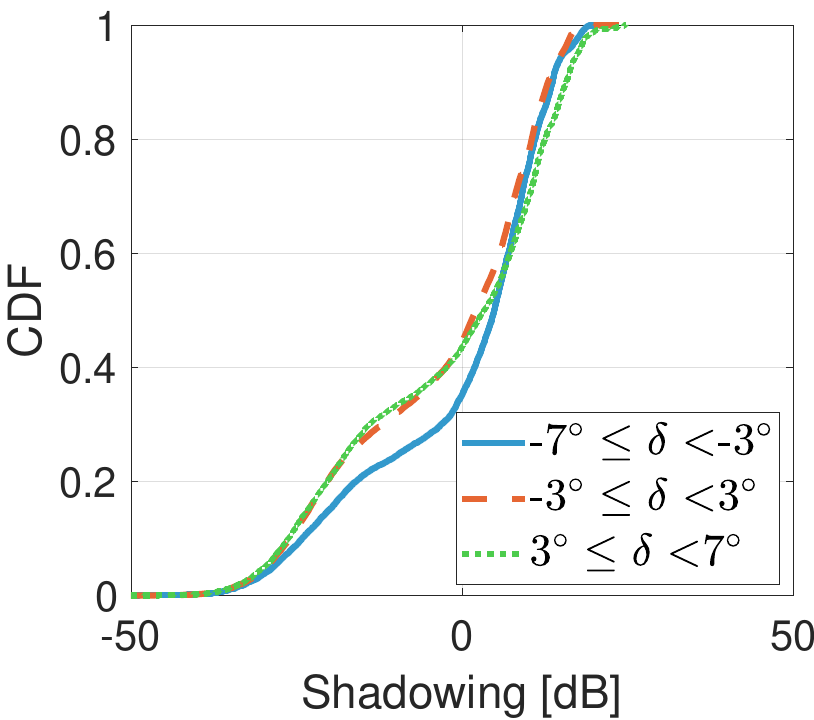}
        \caption{Tilt, \(10^\circ <\theta \leq 30^\circ\)}
        \label{fig:sub1}
    \end{subfigure}
    \vspace{-1mm}
        \caption{CDFs of shadow fading (SF) under different conditioning scenarios: (a) unconditional; (b) conditioned on elevation angle $\theta$; (c) and (d) conditioned jointly on elevation angle $\theta$ and tilt angle $\delta$.} 
    \label{fig:sf_statistics}\vspace{-6mm}
\end{figure}
The full dataset consists of 14 experiments conducted by five university teams. In this work, we use 5 experiments from two North Carolina State University teams, each flown at a UAV altitude of 28~m, as the dataset \(\mathcal{D}'\) for learning the correlation modeling parameters. Cumulative distribution function (CDF) of SF for various tilt and elevation angle scenarios, computed from \(\mathcal{D}'\), are shown in Fig.~\ref{fig:sf_statistics}. We then evaluate Kriging performance on an experiment from the University of North Texas team, conducted at a UAV altitude of 21~m, which is denoted by dataset \(\mathcal{D}\). The test experiment contains approximately 47{,}000 measurement samples. For Kriging, we randomly select \(M\) tuning samples from \(\mathcal{D}\), where \(M\) ranges from 50 to 450, corresponding to roughly \(0.1\%\) to \(1\%\) of the available samples.

Prediction accuracy is quantified using the root mean square error (RMSE) of RSRP. For each value of \(M\), we perform repeated random subsampling experiments: in each trial, a random subset of \(M\) samples is used for training, and a disjoint set of 100 samples is used for testing. This procedure is repeated until a total of 100{,}000 test predictions are obtained. 

\subsection{Angular Correlation Profiles}
To model angular correlations with tilt angle \(\delta\) and elevation angle \(\theta\), we discretize both into a few representative ranges. Specifically, the tilt angle is partitioned into five intervals: (1) \(\delta < -7^\circ\), (2) \(-7^\circ \le \delta < -3^\circ\), (3) \(-3^\circ \le \delta \le 3^\circ\), (4) \(3^\circ < \delta \le 7^\circ\), and (5) \(\delta > 7^\circ\). The corresponding representative (mean) tilt values are chosen as \(-10^\circ\), \(-5^\circ\), \(0^\circ\), \(5^\circ\), and \(10^\circ\), respectively. Similarly, the elevation angle is divided into four ranges: (1) \(0^\circ < \theta \le 10^\circ\), (2) \(10^\circ < \theta \le 30^\circ\), (3) \(30^\circ < \theta \le 50^\circ\), and (4) \(50^\circ < \theta \le 90^\circ\), with representative values \(5^\circ\), \(20^\circ\), \(40^\circ\), and \(70^\circ\), respectively.

\subsubsection{Correlation for Tilt-angle Separation $R_\text{tlt}(\cdot)$}
\begin{figure}[t!]
    \centering
    \begin{subfigure}{0.23\textwidth}
        \centering
        \includegraphics[width=\linewidth,trim={5.6cm 10.0cm 6.2cm 10.0cm},clip]{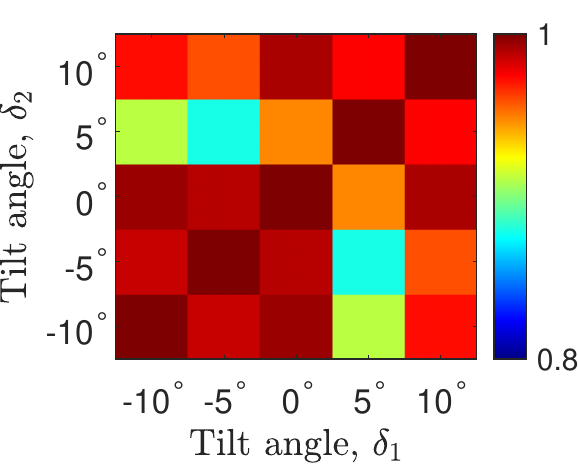}
        \caption{\(\theta \leq 10^\circ\)}
        \label{fig:sub1}
    \end{subfigure}
    \begin{subfigure}{0.23\textwidth}
        \centering
        \includegraphics[width=\linewidth,trim={5.6cm 10.0cm 6.2cm 10.0cm},clip]{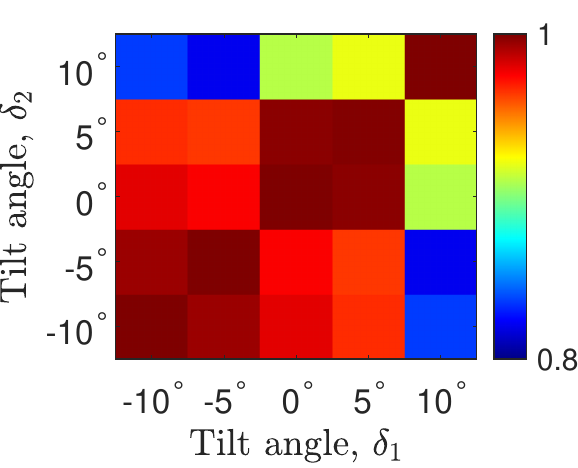}
        \caption{\(10^\circ < \theta \leq 30^\circ\)}
        \label{fig:sub1}
    \end{subfigure}
    \begin{subfigure}{0.23\textwidth}
        \centering
        \includegraphics[width=\linewidth,trim={5.6cm 10.0cm 6.2cm 10.0cm},clip]{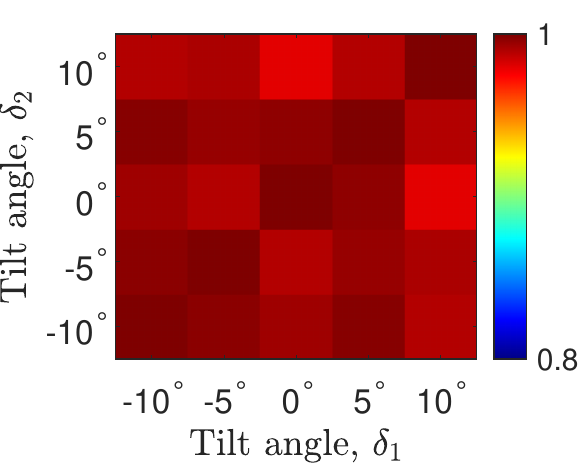}
        \caption{\(30^\circ < \theta \leq 50^\circ\)}
        \label{fig:sub1}
    \end{subfigure}
    \begin{subfigure}{0.23\textwidth}
        \centering
        \includegraphics[width=\linewidth,trim={5.6cm 10.0cm 6.2cm 10.0cm},clip]{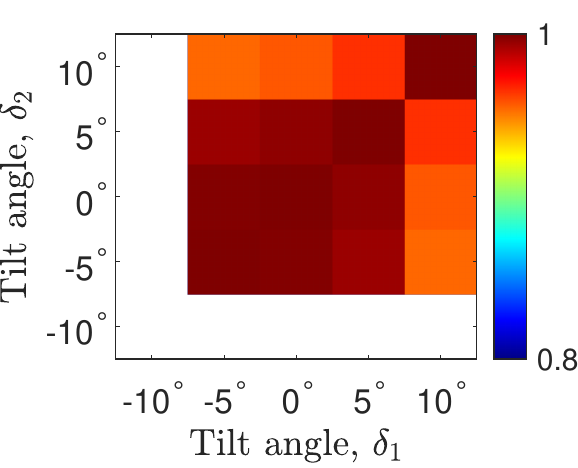}
        \caption{\(50^\circ < \theta \leq 90^\circ\)}
        \label{fig:sub1}
    \end{subfigure}
    \vspace{-1mm}
        \caption{Tilt-angle correlation profiles, conditional on elevation.} 
    \label{fig:emp_tilt_profiles}\vspace{-6mm}
\end{figure}

\begin{figure}[!t]
\centerline{\includegraphics[width=0.9\linewidth,trim={3.0cm 7.5cm 4.3cm 8.1cm},clip]{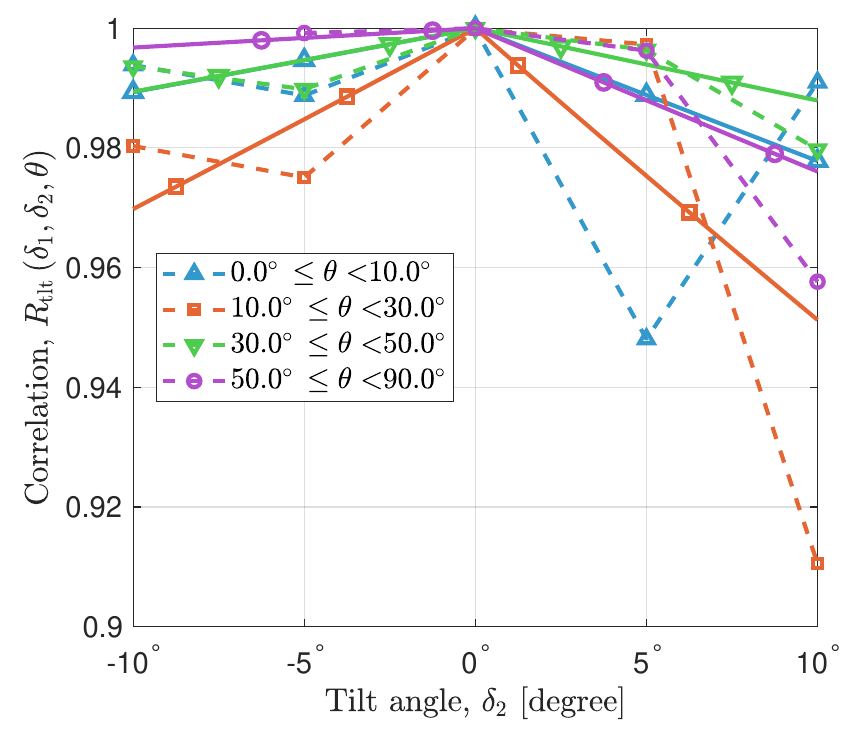}}
\caption{Exponential kernel fitting of empirical correlation values as a function of tilt-angle separation, centered at \(\delta_1 = 0^\circ\). Empirical values are represented in dashed lines, while solid curves represent kernel fits.}
\label{fig:kernel_tilt_profiles}
\vspace{-4mm}
\end{figure}

Fig.~\ref{fig:emp_tilt_profiles} illustrates the empirical tilt-dependent angular correlation profiles \(R_\text{tlt}^{\text{emp}}(\delta_1,\delta_2,\theta)\) for different elevation angle ranges.
At low elevation angles (see Fig.~\ref{fig:emp_tilt_profiles}(a)–(b)), the correlation exhibits a stronger decay with increasing tilt-angle separation, as evidenced by the more pronounced variation in the correlation matrices. This indicates that, at low elevation, changes in tilt angle significantly alter the interaction between the UAV airframe (e.g., fuselage) and the propagation path, thereby impacting SF. In contrast, at intermediate elevation angles (see Fig.~\ref{fig:emp_tilt_profiles}(c)), the correlation matrix is more uniform, suggesting reduced sensitivity to tilt variations. For higher elevation angles (see Fig.~\ref{fig:emp_tilt_profiles}(d)), we observe that the \(\delta \approx 10^\circ\) configuration is particularly sensitive to tilt changes. This can be attributed to the dipole-like antenna pattern: near \(\theta \approx 70^\circ\!-\!80^\circ\), positive tilt angles may steer the main lobe toward \(\theta \approx 90^\circ\), leading to a rapid change in received power. Based on the empirical correlation values, an example of the fitted exponential kernels as a function of tilt-angle separation (see~\eqref{eq:tilt_kernel}) is shown in Fig.~\ref{fig:kernel_tilt_profiles} for a reference tilt angle \(\delta_1 = 0^\circ\). The figure reveals an asymmetric correlation behavior for positive and negative tilt separations, with the fastest decay observed for \(\theta = 20^\circ\) (increasing tilt) and the slowest for \(\theta = 70^\circ\) (decreasing tilt).

\subsubsection{Correlation for Elevation-angle Separation $R_\text{elv}(\cdot)$}
\begin{figure}[t!]
    \centering
    \begin{subfigure}{0.22\textwidth}
        \centering
        \includegraphics[width=\linewidth,trim={5.2cm 10.0cm 6.3cm 10.0cm},clip]{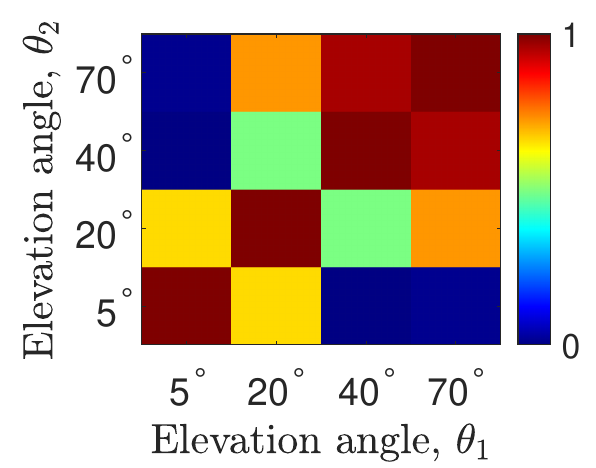}
        \caption{\(-7^\circ <\delta \leq -3^\circ\)}
        \label{fig:sub1}
    \end{subfigure}
    \begin{subfigure}{0.22\textwidth}
        \centering
        \includegraphics[width=\linewidth,trim={5.2cm 10.0cm 6.3cm 10.0cm},clip]{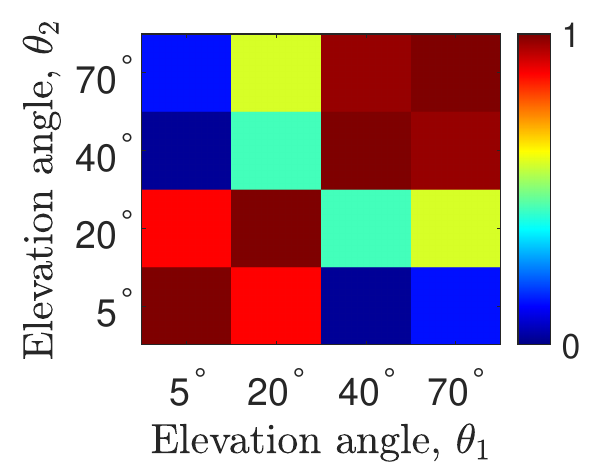}
        \caption{\(-3^\circ <\delta \leq 3^\circ\)}
        \label{fig:sub1}
    \end{subfigure}
    \begin{subfigure}{0.22\textwidth}
        \centering
        \includegraphics[width=\linewidth,trim={5.2cm 10.0cm 6.3cm 10.0cm},clip]{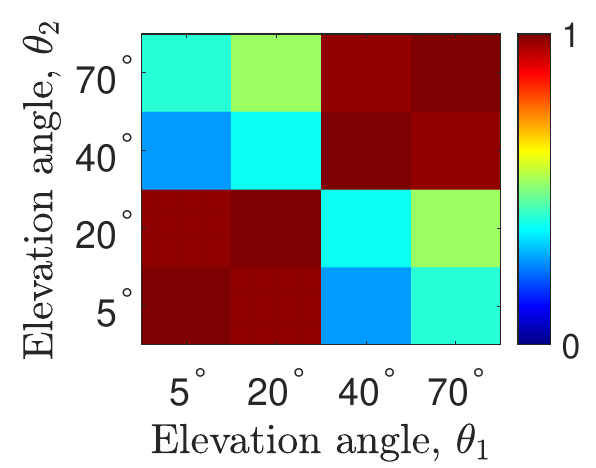}
        \caption{\(3^\circ <\delta \leq 7^\circ\)}
        \label{fig:sub1}
    \end{subfigure}
    \begin{subfigure}{0.22\textwidth}
        \centering
        \includegraphics[width=\linewidth,trim={5.2cm 10.0cm 6.3cm 10.0cm},clip]{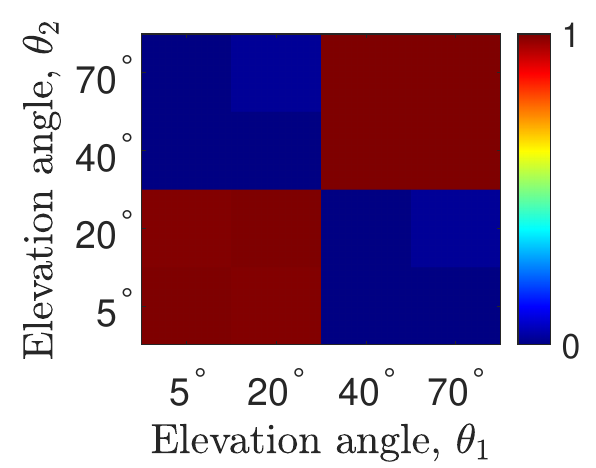}
        \caption{\(\delta >7^\circ\)}
        \label{fig:sub1}
    \end{subfigure}
    \vspace{-1mm}
        \caption{Elevation-angle correlation profiles, conditional on tilt.} 
    \label{fig:emp_elev_profiles}\vspace{-6mm}
\end{figure}

\begin{figure}[!t]
\centerline{\includegraphics[width=0.9\linewidth,trim={3.0cm 7.5cm 4.3cm 8.1cm},clip]{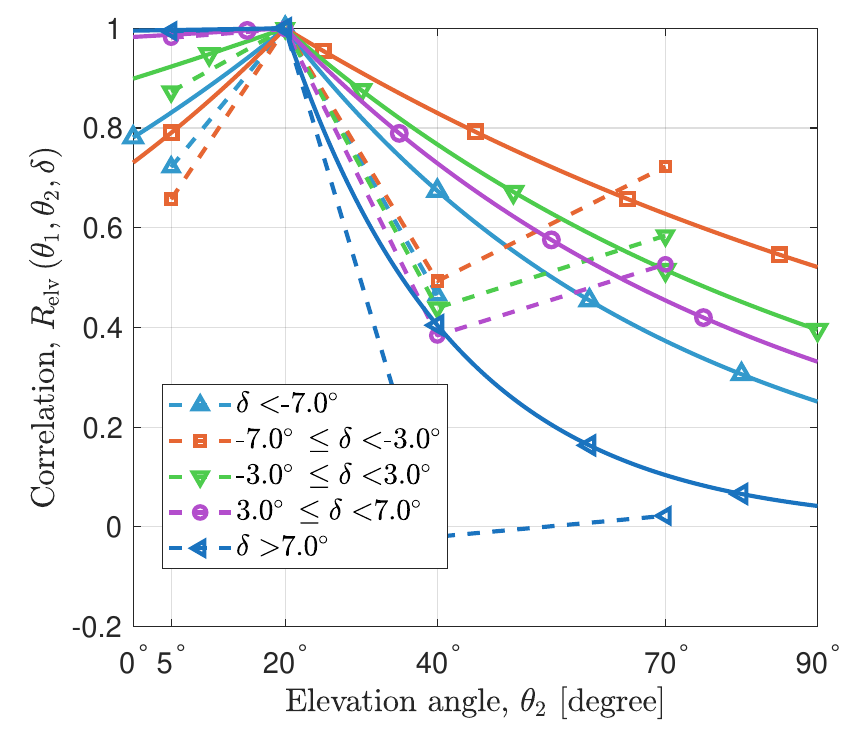}}
\caption{Exponential kernel fitting of empirical correlation values as a function of elevation-angle separation, centered at \(\theta_1 = 20^\circ\). Empirical values are represented in dashed lines, while solid curves represent kernel fits.}
\label{fig:kernel_elev_profiles}
\vspace{-4mm}
\end{figure}

Fig.~\ref{fig:emp_elev_profiles} shows the empirical elevation-dependent angular correlation profiles \(R_\text{elv}^{\text{emp}}(\theta_1,\theta_2,\delta)\) for different tilt-angle ranges.
For negative tilt angles (see Fig.~\ref{fig:emp_elev_profiles}(a)), which increase the interaction between the UAV fuselage and the LoS path, the correlation decays rapidly as the elevation changes from \(5^\circ\) to \(20^\circ\). In contrast, for zero and positive tilt angles (see Fig.~\ref{fig:emp_elev_profiles}(b)–(d)), the correlation decays more slowly when moving from \(5^\circ\) to \(20^\circ\). This is consistent with the intuition that, as the elevation increases, the relative impact of a negative tilt angle on fuselage-induced shadowing is reduced.
Across all tilt-angle ranges, a pronounced change in correlation is observed when the elevation increases from \(20^\circ\) to \(40^\circ\); beyond this point, the correlation remains relatively high between \(40^\circ\) and \(70^\circ\). This behavior is particularly evident for larger positive tilt angles in Fig.~\ref{fig:emp_elev_profiles}(d). These trends suggest that, under clear LoS conditions, SF statistics undergo a major transition between \(20^\circ\) and \(40^\circ\), indicating that shadowing is strongly governed by elevation in this dataset.
This is further supported by Fig.~\ref{fig:kernel_elev_profiles}, which shows an example of exponential-kernel fitting to the empirical correlation values as a function of elevation separation (see~\eqref{eq:elev_kernel}). The fitted curves show that, for increasing elevation beyond \(20^\circ\), the correlation decays sharply, with the steepest decay observed at more positive \(\delta\), corresponding to a clear LoS scenario.

\subsection{Kriging Interpolation Accuracy}
OK critically depends on accurate correlation estimation among measured data points. In the conventional approach, this correlation is modeled purely as a function of spatial distance, as in~\eqref{eq:dedm}. In contrast, the proposed correlation model in~\eqref{eq:proposed_full_r} augments the distance-based term by incorporating both tilt and elevation angles. In this subsection, we compare these two correlation formulations in terms of their impact on Kriging interpolation performance. 

\subsubsection{Performance Summary in Terms of Median RMSE}
\begin{figure}[t!]
    \centering
    \begin{subfigure}{0.38\textwidth}
        \centering
        \includegraphics[width=\linewidth,trim={0.75cm 5.35cm 1.55cm 6.3cm},clip]{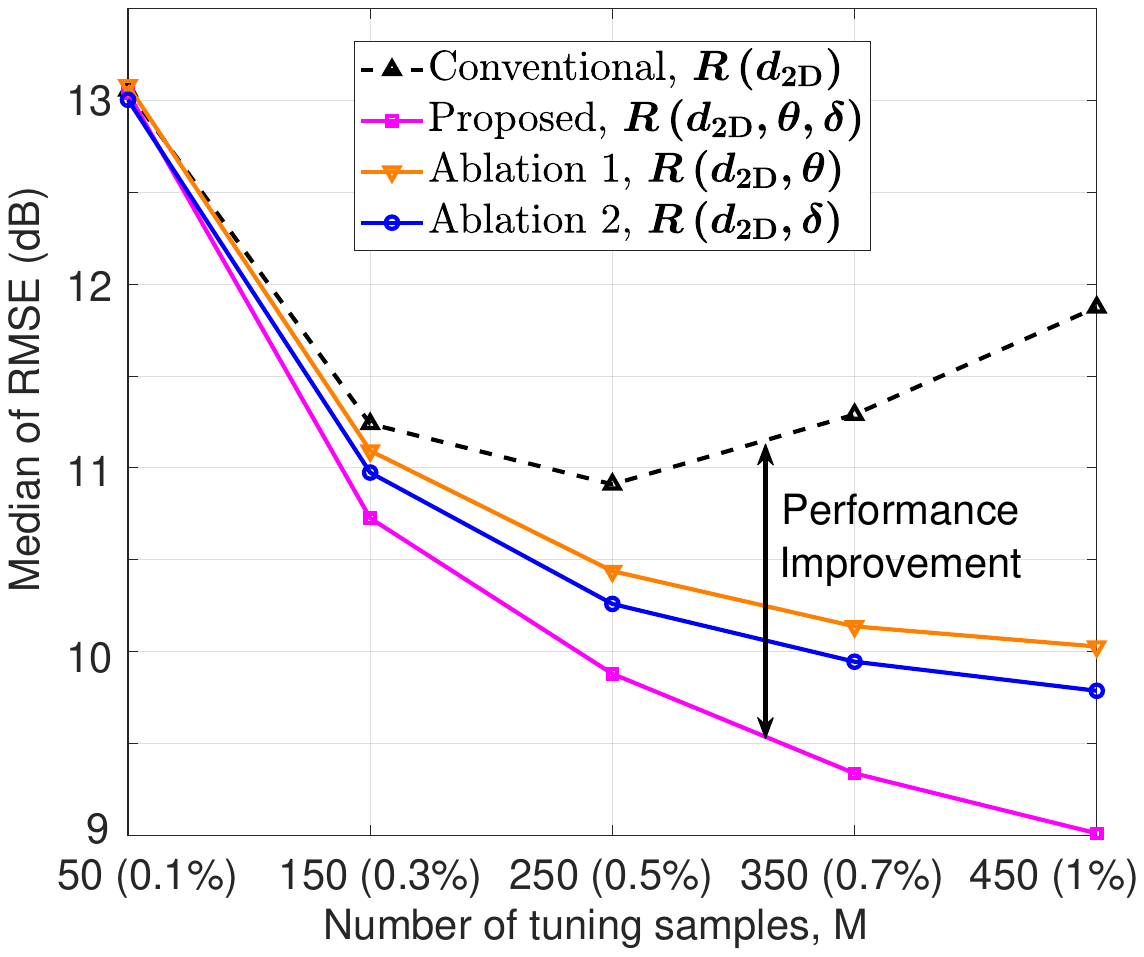}
        \caption{Median}
        \label{fig:sub1}
    \end{subfigure}
    \begin{subfigure}{0.45\textwidth}
        \centering
        \includegraphics[width=\linewidth,trim={2.3cm 7.0cm 2.65cm 7.5cm},clip]{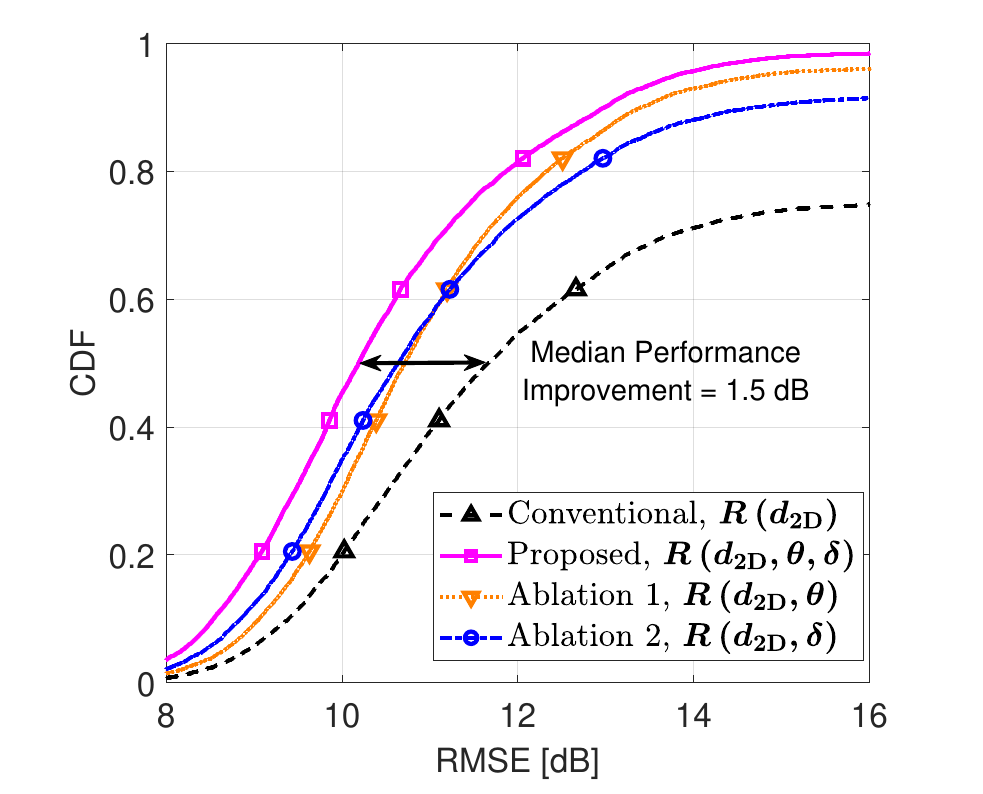}
        \caption{CDF}
        \label{fig:sub1}
    \end{subfigure}
        \caption{Kriging performance comparison with and without angle-dependent correlation modeling: (a) median RMSE plot for varying number of tuning samples, (b) RMSE CDFs.}
    \label{fig:rmse_301}\vspace{-6mm}
\end{figure}
Fig.~\ref{fig:rmse_301}(a) compares the median RMSE for the conventional and proposed correlation models.
The proposed model yields an improvement of approximately 2~dB at \(M = 350\) (i.e., about \(0.7\%\) of the available samples). Moreover, the performance gain increases with the number of tuning samples. This is noteworthy because Kriging often struggles as the number of training samples increases, due to the growing size of the covariance matrix and potential numerical issues. In contrast, with the proposed angle-aware correlation model, the median RMSE continues to decrease as \(N\) increases. This suggests that the enriched correlation structure mitigates issues such as near-singular covariance matrices and negative Kriging weights that can destabilize predictions under a purely distance-based model.
To further examine the prediction behavior, Fig.~\ref{fig:rmse_301}(b) shows the CDFs of the RMSE values.
The proposed method substantially reduces the occurrence of large-error (``bad'') predictions while also improving overall accuracy, confirming the benefits of incorporating angular information. These results indicate that the proposed correlation model, which jointly accounts for distance, tilt, and elevation, provides a more faithful representation of the underlying SF process. To assess the importance of joint modeling, we also perform two ablation studies in which only a single angular variable (tilt-only or elevation-only) is used to construct the angular correlation. In both cases, the performance is inferior to the proposed joint model, demonstrating that capturing the combined effects of tilt and elevation is critical for achieving the observed gains.

\subsubsection{Comparison of Reconstructed SF Maps}
\begin{figure}[t!]
    \centering
    \begin{subfigure}{0.23\textwidth}
        \centering
        \includegraphics[width=\linewidth,trim={3.4cm 8.5cm 4.2cm 8.35cm},clip]{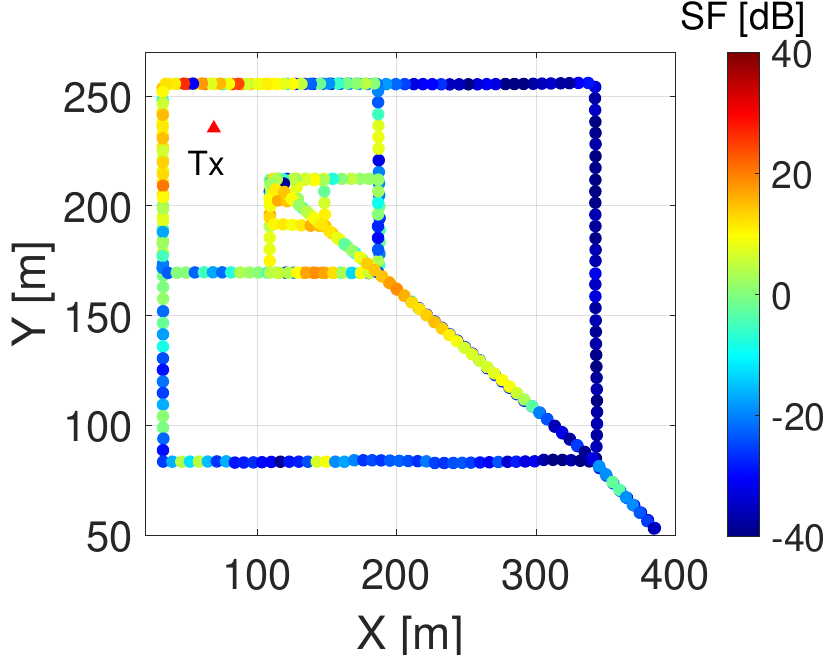}
        \caption{All measurements}
        \label{fig:sub1}
    \end{subfigure}
    \begin{subfigure}{0.23\textwidth}
        \centering
        \includegraphics[width=\linewidth,trim={3.4cm 8.5cm 4.2cm 8.35cm},clip]{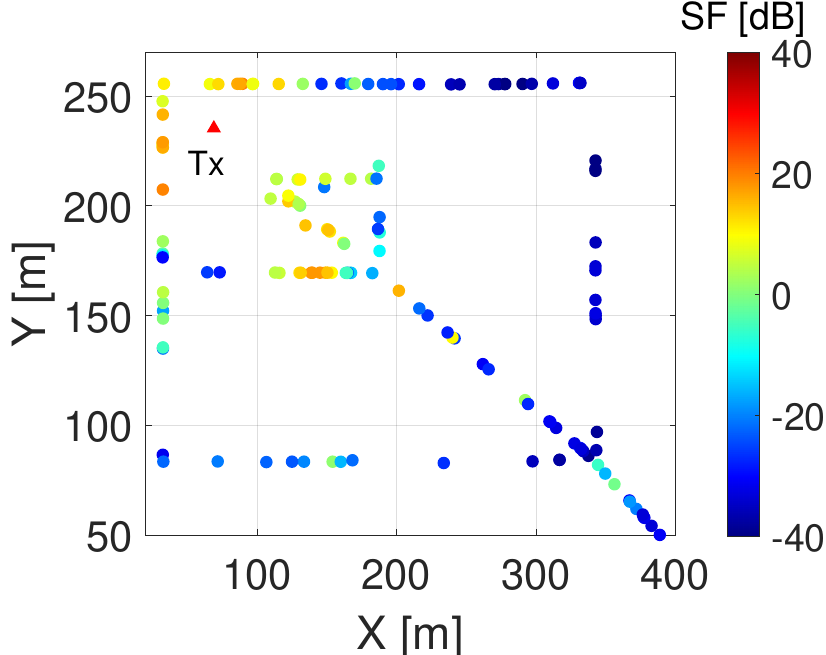}
        \caption{Tuning samples (M = 150)}
        \label{fig:sub1}
    \end{subfigure}
    \begin{subfigure}{0.23\textwidth}
        \centering
        \includegraphics[width=\linewidth,trim={3.4cm 8.5cm 4.2cm 8.2cm},clip]{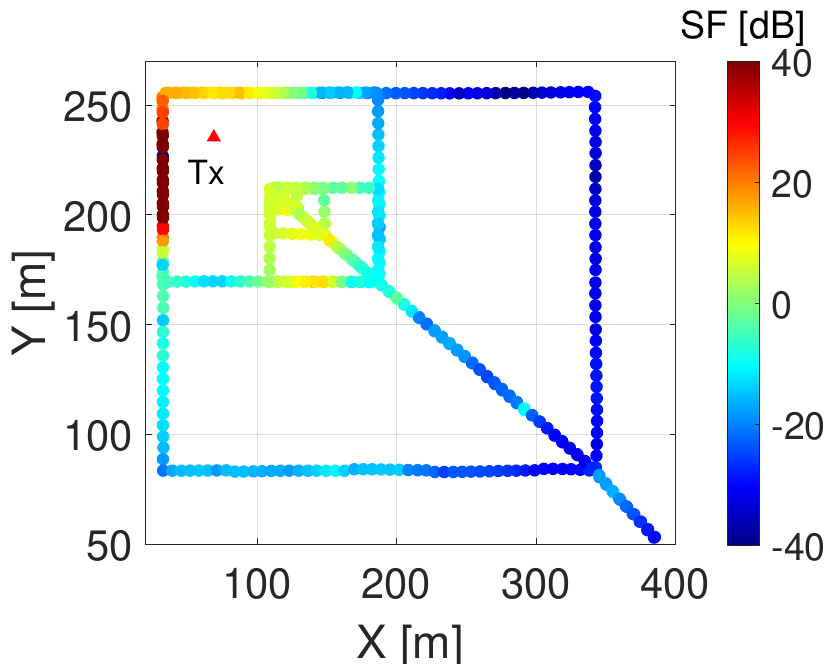}
        \caption{OK (conv.)}
        \label{fig:sub1}
    \end{subfigure}
    \begin{subfigure}{0.23\textwidth}
        \centering
        \includegraphics[width=\linewidth,trim={3.4cm 8.5cm 4.2cm 8.2cm},clip]{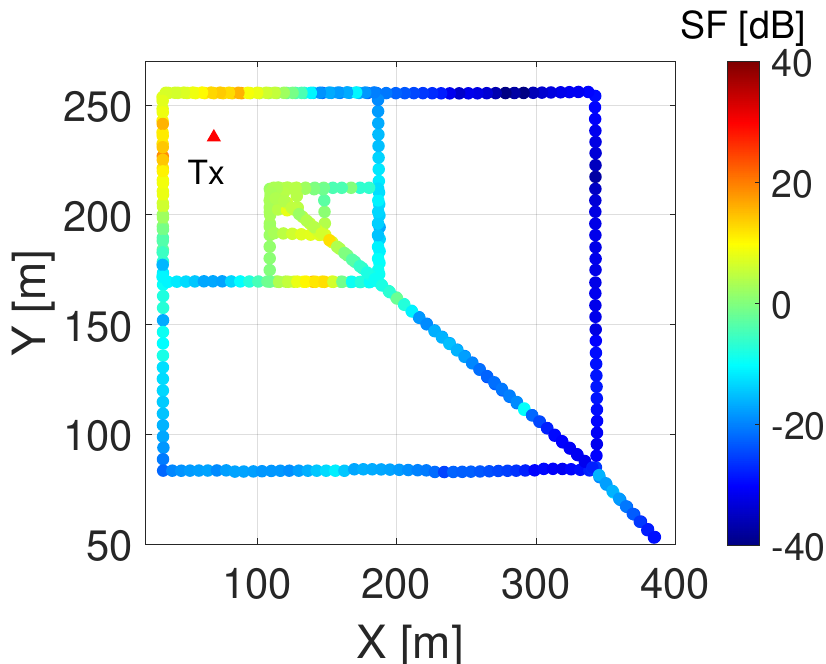}
        \caption{OK w/ ang. corr. (prop.)}
        \label{fig:sub1}
    \end{subfigure}
    \vspace{-1mm}
        \caption{Shadow fading (SF) maps and interpolation results on dataset \(\mathcal{D}\): (a) SF values computed from all measurement samples; (b) randomly selected training samples (\(M=150\)); (c) SF prediction using ordinary Kriging (OK) with conventional distance-based correlation; (d) SF prediction using the proposed method with elevation- and tilt-dependent correlation.}
    \label{fig:pictorial_kriging}\vspace{-6mm}
\end{figure}

To gain further insight into how the angle-aware correlation model improves Kriging performance, we compare the reconstructed SF maps obtained with the conventional and proposed approaches in Fig.~\ref{fig:pictorial_kriging}.
The key difference between the conventional map in Fig.~\ref{fig:pictorial_kriging}(c) and the proposed map in Fig.~\ref{fig:pictorial_kriging}(d) appears in the upper-left portion of the UAV trajectory, where the conventional approach predicts unrealistically large SF values. As shown in Fig.~\ref{fig:pictorial_kriging}(b), there are several training samples located in this region at comparable distances from the prediction points. Under the conventional, purely distance-based correlation model, these nearby samples effectively compete with one another, which can lead to unstable predictions and exaggerated SF estimates.
In contrast, with the proposed angle-aware correlation model, the variation of elevation angle along the trajectory, particularly pronounced due to the proximity of the transmitter, is explicitly taken into account. In this region, small changes in elevation angle induce non-negligible changes in SF, which a distance-only model is unable to capture. The angle-aware formulation, on the other hand, assigns a higher weight to tuning samples that are not only spatially close but also similar in elevation (and tilt) to the prediction point. As a result, the reconstructed SF map in Fig.~\ref{fig:pictorial_kriging}(d) is more consistent.

\section{Conclusion}
\label{sec:conclution}
In this study, we have investigated the impact of incorporating elevation and tilt angles into SF correlation modeling. Our results demonstrate that SF behavior is jointly governed by both angles, implicitly capturing airframe shadowing, the dipole-like antenna pattern, and elevation-dependent SF. The impact of tilt is most pronounced near \(0^\circ\) and \(90^\circ\) elevation angles.
 The resulting angular correlation exhibits asymmetric decay with respect to increasing versus decreasing angles.
The proposed angle-aware correlation model, when integrated into the Kriging framework, significantly reduces the incidence of large-error “bad” predictions and continues to yield performance gains as the number of tuning samples increases, whereas the traditional distance-based correlation model tends to degrade or saturate with a large number of training points. 
Future work will explore alternative kernel families beyond the exponential form, extend the analysis to multi-UAV measurement campaigns, and incorporate statistical confidence measures for the learned kernels.

\bibliographystyle{IEEEtran}
\bibliography{ref}

\end{document}